\documentclass[twocolumn,longbibliography]{revtex4-1}
\usepackage{times}
\usepackage{amsmath}
\usepackage{amsfonts}
\usepackage{graphicx}% Include figure files
\usepackage{hyperref}
\usepackage{dcolumn}% Align table columns on decimal point
\usepackage{bm}% bold math
\usepackage{color}
\usepackage{xcolor}
\usepackage{soul} % do not alter references as ulem does, but one cannot do \rt{ \ref{} }
\usepackage{orcidlink}
\def\beq{\begin{equation}}
\def\eeq{\end{equation}}
\def\bea{\begin{eqnarray}}
\def\eea{\end{eqnarray}}
\def\beqn{\begin{eqnarray}} 
\def\eeqn{\end{eqnarray}}
\def\beeq{\begin{eqnarray}}
\def\eeeq{\end{eqnarray}}

\def\nn{\nonumber}
\def\Eq#1{Eq.~(\ref{#1})}

\def\qon#1{q_{#1,0}^{(+)}}

\def\qb{\mathbf{q}}

\def\lb{\boldsymbol{\ell}}

\def\ii{\imath 0}

\def\ad#1{{\cal A}_{\rm D}^{(#1)}}

\def\ps#1{\widetilde \Delta_{#1}}

\begin{document}

%%%%%%%%%%%%%%%%%%%%%%%%%%%%%%%%%%%%%%%%%%%%%%%%%%%%%%%%%%%%%%%%%%%%%
%\preprint{IFIC/XX-XX}

%\title{Quantum integration of vacuum amplitudes and time-like causal unitary in the loop-tree duality}
%\title{Quantum integration of decay rates from causal unitary in the loop-tree duality theory}
\title{Quantum integration of decay rates  at second order in perturbation theory}

\author{Jorge J. Mart\'{\i}nez de Lejarza\orcidlink{0000-0002-3866-3825}$^a$}\email[Corresponding author: ]{jormard@ific.uv.es}
\author{David F. Renter\'{\i}a-Estrada\orcidlink{0000-0002-2825-9837}~$^{a}$}%\email{david.renteria@ific.uv.es}
\author{Michele Grossi\orcidlink{0000-0003-1718-1314}$^b$}%\email{michele.grossi@cern.ch}
\author{Germán Rodrigo\orcidlink{0000-0003-0451-0529}$^a$}%\email{german.rodrigo@csic.es}%
\affiliation{%
 $^a$Instituto de F\'{\i}sica Corpuscular, Universitat de Val\`encia - Consejo Superior de Investigaciones Cient\'{\i}ficas, Parc Cient\'{\i}fic, E-46980 Paterna, Valencia, Spain
}%
\affiliation{%
$^b$European Organization for Nuclear Research (CERN), 1211 Geneva, Switzerland 
}%

\date{\today}

\begin{abstract}
We present the first quantum computation of a total decay rate in high-energy physics at second order in perturbative quantum field theory. This work underscores the confluence of two recent cutting-edge advances. On the one hand, the quantum integration algorithm Quantum Fourier Iterative Amplitude Estimation
(QFIAE), which efficiently decomposes the target function into its Fourier series through a quantum neural network before quantumly integrating the corresponding Fourier components. On the other hand, causal unitary in the loop-tree duality (LTD), which exploits the causal properties of vacuum amplitudes in LTD to coherently generate all contributions with different numbers of final-state particles to a scattering or decay process, leading to singularity-free integrands that are well suited for Fourier decomposition. We test the performance of the quantum algorithm with benchmark decay rates in a quantum simulator and in quantum hardware, and find accurate theoretical predictions in both settings.
\end{abstract}

\maketitle

%%%%%%%%%%%%%%%%%%%%%%%%%%%%%%%%%%%%%%%%%%%%%%%%%%%
\section{Introduction}
\label{sec:intro}

The interplay between high-energy physics and quantum computing represents a promising frontier for advancing our understanding of fundamental concepts and improving computational techniques.  High-energy physics requires complex theoretical calculations to predict with high accuracy cross sections and decay rates, which are essential for understanding the behavior of elementary particles at quantum scales and for validating theoretical models in quantum field theory~\cite{Heinrich:2020ybq}. These complex calculations often challenge classical computational methods. Quantum computing, with its inherent ability to leverage the principles of quantum mechanics, offers a novel approach to successfully addressing these challenges~\cite{DiMeglio:2023nsa,Delgado:2022tpc,Rodrigo:2024say}. The applications of quantum computing in this field include jet identification and clustering~\cite{thaler,delgado_jets,lejarza,deLejarza:2022vhe,Gianelle:2022unu}, parton density determination and integration~\cite{carrazza,Cruz-Martinez:2023vgs}, parton shower simulation~\cite{spannowsky}, anomaly detection~\cite{Belis:2023atb,Schuhmacher:2023pro,Bermot:2023kvh,Ngairangbam:2021yma}, integration of elementary particle process ~\cite{AGLIARDI2022137228}, data classification~\cite{
Belis:2024guf,Belis:2021zqi, Blance:2020ktp,Heredge:2021vww,Peixoto:2022zzk,Hammad:2023wme,Lazar:2024luq} and the analysis of the causal structure of multiloop Feynman diagrams~\cite{Ramirez-Uribe:2021ubp,Clemente:2022nll}.  

In particle physics, Feynman diagrams and scattering amplitudes from perturbative quantum field theory are essential tools for predicting the transition probabilities between particle states at high-energy colliders, such as the CERN's Large Hadron Collider. Loop Feynman diagrams represent interactions between particles that involve virtual quantum fluctuations, making them inherently complex. On the other hand, tree-level Feynman diagrams represent direct interactions between particles, and although they are apparently easier to evaluate, they are not exempt from difficulties. Traditional methods for evaluating Feynman diagrams and scattering amplitudes, and combining them to extract accurate theoretical predictions, while effective, are limited by their computational complexity and the resources required to perform, for example, numerical integrations over the loop momenta and the phase space of the final states. 

The loop-tree duality (LTD) framework~\cite{Catani:2008xa,Bierenbaum:2010cy,Bierenbaum:2012th,Buchta:2014dfa,Buchta:2015wna,Capatti:2019edf} facilitates the evaluation of multiloop Feynman diagrams by decomposing them into tree-like objects, providing a structured approach to these computations where the fundamental physical principle of causality is manifest in the integrand representation~\cite{Aguilera-Verdugo:2020set,Aguilera-Verdugo:2020kzc,Ramirez-Uribe:2020hes,JesusAguilera-Verdugo:2020fsn,Ramirez-Uribe:2022sja,Sborlini:2021owe,TorresBobadilla:2021ivx,Aguilera-Verdugo:2019kbz}. We have recently proposed a novel approach based on LTD to efficiently recast perturbative theoretical predictions at high-energy colliders, the LTD causal unitary~\cite{Ramirez-Uribe:2024rjg,LTD:2024yrb}, where differential observables, cross sections and decay rates are assembled from the LTD representation of vacuum amplitudes, i.e. scattering amplitudes without external particles. 

In Ref.~\cite{deLejarza:2023qxk,deLejarza:2023IEEE}, we have also proposed a quantum integration algorithm, dubbed Quantum Fourier Iterative Amplitude Estimation (QFIAE), and in Ref.~\cite{deLejarza:2024pgk} we have applied this quantum integrator to the evaluation of infrared-safe scalar one-loop Feynman integrals. 

%\mg{However, the integration processes involved in this framework are computationally intensive, often necessitating sophisticated techniques to handle the causal unitary elements within. Quantum algorithms can potentially reduce the complexity of integration tasks, improving both the speed and accuracy of results.  \\ }

By integrating quantum computing techniques into the LTD framework and harnessing the power of quantum integration, we expect a transformative approach that could lead to new insights and more efficient methodologies. Therefore, the aim of this work is going one step further and test the performance of QFIAE with physical decay rates at second order in perturbation theory or next-to-leading order (NLO). This requires the combination of one-loop with tree-level contributions, where each of the contributions is individually singular and therefore numerically challenging, although the final prediction is finite. 
We base our approach on the LTD causal unitary framework because the unified treatment of loop and tree-level contributions leads to rather flat integrands, and therefore integrands that are more suitable for numerical integration, in particular, by Fourier decomposition.

%%%%%%%%%%%%%%%%%%%%%%%%%%%%%%%%%%%%%%%%%%%%%%%%%%%%%%%%%%
\section{Quantum Fourier Iterative Amplitude Estimation}
\label{sec:qfiae}
%%%%%%%%%%%%%%%%%%%%%%%%%%%%%%%%%%%%%%%%%%%%%%%%%%%%%%%%%%

\begin{figure}[t]
\begin{center}
\includegraphics[scale=0.245]{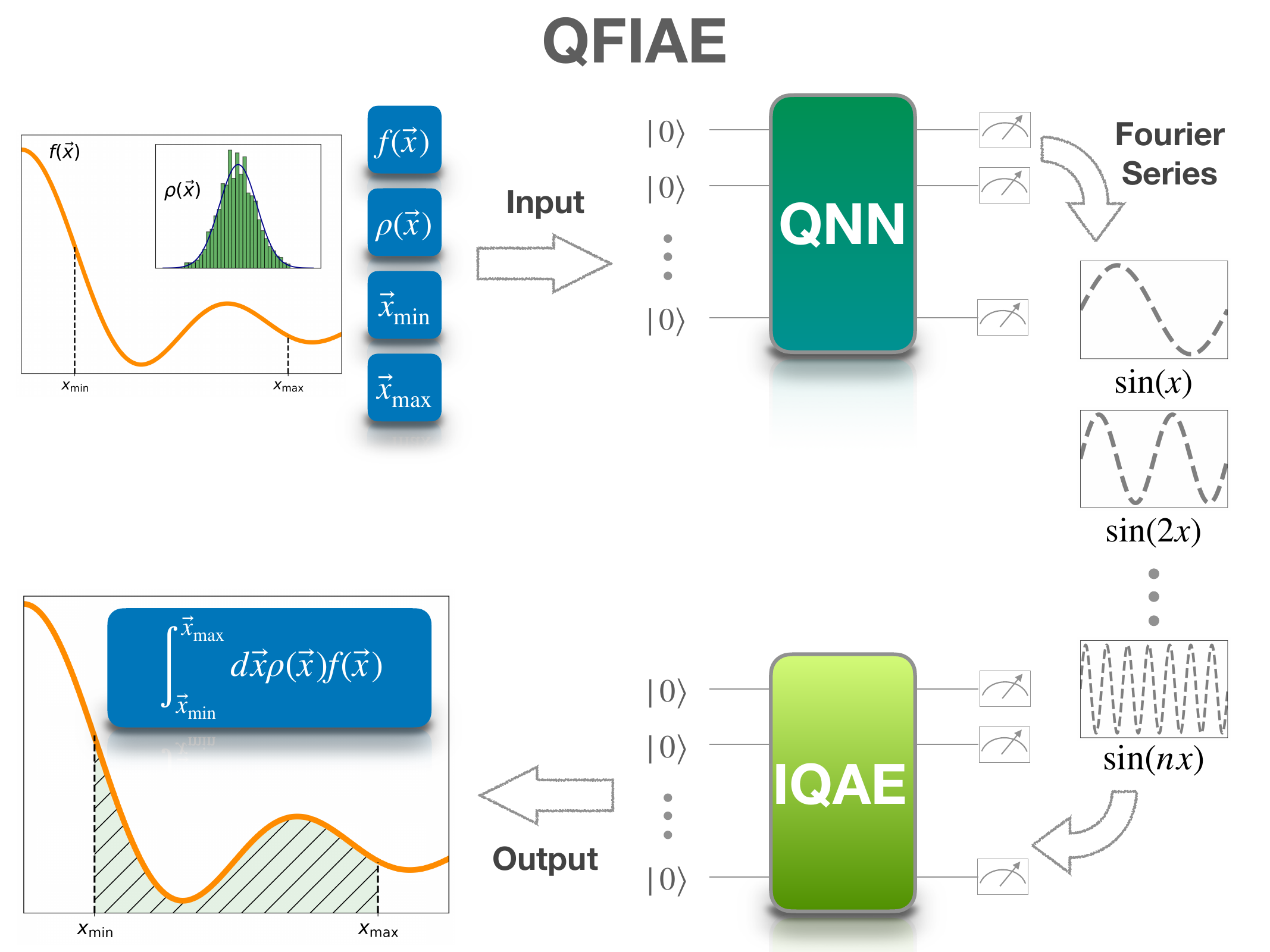}
\caption{Workflow of QFIAE. The input consists of the target function $f(\vec{x})$, the probability distribution $\rho(\vec{x})$, and the integration domain $\{\vec{x}_{min},\vec{x}_{max} \}$. The QNN fits $f(\vec{x})$ and extracts its Fourier series from the quantum circuit. Next, IQAE estimates the integral for each trigonometric term in the Fourier series. Finally, these integrals are added with their corresponding coefficients to obtain the final integral result. 
\label{fig:sketch qfiae}}
\end{center}
\end{figure}

Quantum Fourier Iterative Amplitude Estimation~(QFIAE)~\cite{deLejarza:2023qxk,deLejarza:2023IEEE,deLejarza:2024pgk} is a quantum algorithm designed to efficiently integrate multidimensional functions. Its workflow is depicted in Fig.~\ref{fig:sketch qfiae}. QFIAE first decomposes the target function into its Fourier series using a Quantum Neural Network~(QNN) via a data re-uploading approach~\cite{Schuld_2021,Casas:2023ure,P_rez_Salinas_2020}. Previous studies~\cite{Schuld_2021,gil2020input} have demonstrated that an exponential data encoding results in the quantum model representing a truncated Fourier series. Following this first step, each trigonometric term of the Fourier series undergoes quantum integration using Iterative Quantum Amplitude Estimation~(IQAE)~\cite{Grinko_2021}, which is an efficient variant of Quantum Amplitude Estimation (QAE)~\cite{Brassard_2002}.

%\rd{comment on the probability distribution $\rho(\vec x)$}

%%%%%%%%%%%%%%%%%%%%%%%%%%%%%%%%%%%%%%%%%%%%%%%%%%%%%%%%%%
%%%%%%%%%%%%%%%%%%%%%%%%%%%%%%
%%%%%%%%%%%%%%%%%%%%%%%%%%%%%%%%%%%%%%%%%%%%%%%%%%%%%%%%%%

The Fourier decomposition enables encoding the target function with minimal quantum arithmetic operations and also capitalizes on the quantum-friendly nature of the sine function for integration purposes. At the core of QFIAE lies the QNN, which provides a practical approach to preserving the potential quadratic speedup in the number of queries to the probability distribution function, which will be encoded into the amplification operator, %\rd{used to .... }, 
offered by the Amplitude Amplification algorithm underlying in QAE, compared to other recently proposed quantum integration algorithms like Fourier Quantum Monte Carlo Integration~(FQMCI)~\cite{Herbert_2022}. FQMCI, also employs Fourier decomposition to approximate the integrand and then individually estimate each component using QAE. However, FQMCI relies on assumptions about acquiring the Fourier coefficients, which may not always hold. Failure to meet these assumptions can nullify the potential quantum speedup. The QNN, on the other hand, ensures a reliable quantum extraction of the Fourier coefficients. A detailed comparison of the performance of both FQMCI and QFIAE can be found in \cite{deLejarza:2023IEEE}.

The second critical aspect of QFIAE involves exploiting the advantages of IQAE over QAE. QAE estimates quantum state amplitudes using amplitude amplification, an extension of Grover's algorithm~\cite{Grover:1997fa}, to enhance the likelihood of measuring the desired state over undesired states. However, QAE is constrained by its dependence on the resource-intensive Quantum Phase Estimation (QPE) subroutine~\cite{9781107002173}, which entails operations deemed computationally expensive for current Noisy Intermediate Scale Quantum (NISQ) devices. This limitation threatens the anticipated quadratic advantage promised by QAE. IQAE addresses this challenge by replacing QPE with a classically efficient post-processing method, reducing the demands on qubits and quantum gates while preserving the asymptotic quadratic speedup.

%%%%%%%%%%%%%%%%%%%%%%%%%%%%%%%%%%%%%%%%%%%%%%%%%%%%%%%%%%%%%%%%%%
\section{LTD causal unitary and decay rates at NLO}\label{sec:ltd}

A vacuum amplitude in LTD, $\ad{\Lambda}$, where $\Lambda$ is the number of primitive loop four-momenta, is obtained from its Feynman representation by integrating out through the Cauchy residue theorem one component of each primitive loop momenta~\cite{Catani:2008xa,Bierenbaum:2010cy}, typically the energy components, which results in replacing the Feynman propagators by causal propagators of the form~\cite{Aguilera-Verdugo:2020set}
\beq
\frac{1}{\lambda_{i_1 \cdots i_m}} = \left(\sum_{s=1}^m \qon{i_s} \right)^{-1}~,
\label{eq:causalvacuum}
\eeq
where $\qon{i_s} =\sqrt{\qb_{i_s}^2+m_{i_s}^2-\ii}$ are the on-shell energies of the internal propagators, with $\qb_{i_s}$ the spatial components of the four-momenta and $m_{i_s}$ their masses. The numerator of the vacuum amplitude in LTD is also a function of the on-shell energies and additionally of the internal masses. The factor $\imath 0$ in the on-shell energies stems from the original infinitesimal complex prescription of the Feynman propagators. Loop vacuum amplitudes in the customary Feynman representations are functions in the Minkowski space of the loop four-momenta, while the integration domain of loop vacuum amplitudes in LTD is the Euclidean space of the loop three-momenta.

Each causal propagator, \Eq{eq:causalvacuum}, involves a set of internal particles that divide the vacuum amplitude into two subamplitudes, with the momentum flow of all particles in the set aligned in the same direction, and each term in the vacuum amplitude is proportional to a product of causal propagators in which the momentum flow of the shared particles are also aligned in the same direction~\cite{Aguilera-Verdugo:2020kzc,Ramirez-Uribe:2020hes,JesusAguilera-Verdugo:2020fsn,Ramirez-Uribe:2022sja,Sborlini:2021owe,TorresBobadilla:2021ivx}. This picture is also analogous to selecting the acyclic configurations of a directed graph in graph theory~\cite{Ramirez-Uribe:2021ubp,Clemente:2022nll,Ramirez-Uribe:2024wua}. In the limit where a causal propagator becomes singular, all the particles involved are set on shell. Therefore a natural procedure to generate all the interferences of scattering amplitudes with different numbers of final-state particles and loops that are considered in the state-of-the-art approaches is to take residues on the causal propagators. This is the central idea of LTD causal unitary~\cite{Ramirez-Uribe:2024rjg,LTD:2024yrb}. The vacuum amplitude in LTD thus acts as a kernel amplitude, which generates all the final states contributing to a scattering or decay process from all possible residues on causal propagators. 

\begin{figure}[t]
\begin{center}
\includegraphics[scale=0.38]{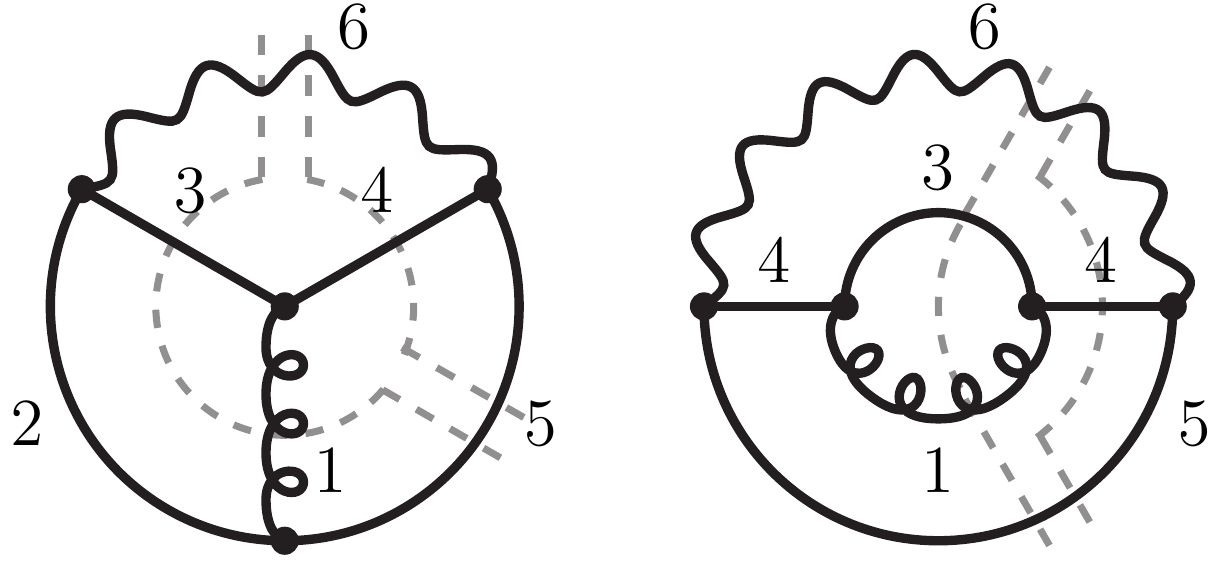}
\caption{Three-loop vacuum diagrams contributing to the decay $\gamma^*\to q \bar q (g)$ at NLO. The gray dashed lines represent phase-space residues, i.e. different final states. Similar diagrams contribute to the decays $H\to q\bar q (g)$ and $\Phi\to \phi\phi (\phi)$ by substituting the photon labeled~$6$ by a Higgs boson or a heavy scalar $\Phi$; for the heavy scalar decay particles~$1$ to $5$ are substituted by light scalars.
\label{fig:qqbar}}
\end{center}
\end{figure}

As benchmark decay rates at NLO, we consider the decay of a heavy scalar into lighter scalars, and the decay of a Higgs boson or an off-shell photon into a pair of massive quarks and antiquarks. These processes have been implemented for a proof of concept of LTD causal unitary in Ref.~\cite{LTD:2024yrb}, where classical integration methods were used to predict the total decay rates. We refer to Ref.~\cite{LTD:2024yrb} for a detailed presentation of the expressions used in the numerical implementation. The vacuum diagrams that contribute to the decay $\gamma^* \to q\bar q (g)$ are shown in Fig.~\ref{fig:qqbar}. Similar vacuum diagrams describe the other two decay processes considered.

%%%%%%%%%%%%%%%%%%%%%%%%%%%%%%
%%%%%%%%%%%%%%%%%%%%%%%%%%%%%%

The vacuum diagrams in Fig.~\ref{fig:qqbar} contribute to a vacuum amplitude that in LTD depends on three loop three-momenta, $\{\lb_1, \lb_2, \lb_3\}$. The three-momenta of the internal propagators read
\bea
\qb_1 = \lb_1+\lb_2~, \quad 
\qb_2 = \lb_1+\lb_3~, \quad
\qb_3 = \lb_1~, \nn \\ 
\qb_4 = \lb_2~, \quad 
\qb_5 = \lb_2-\lb_3~, \quad
\qb_6 = \lb_3~, 
\eea
and the corresponding on-shell energies are given by $\qon{i}  = \sqrt{\qb_i^2+m_i^2-\ii}$. We work in the rest frame of the decaying particle, where $\lb_3=\boldsymbol{0}$. Therefore, the unintegrated decay rate is a function of the remaining three-momenta, $\lb_1$ and $\lb_2$, through the on-shell energies.

The differential decay rate of a particle $a$ at NLO takes the form
\bea
d\Gamma^{(1)}_{a} &=& \frac{d\Phi_{\lb_1\lb_2}}{2\sqrt{s}} \, \bigg[
\Big( \ad{3,a}(456) \, \ps{45\bar 6}  + \ad{3,a}(1356) \, \ps{135\bar 6} \Big) \nn \\ &+& (5\leftrightarrow 2, 4\leftrightarrow 3) \bigg]~,
\label{eq:decayratescalarNLO}
\eea
where $\ad{3,a}(456)$ and $\ad{3,a}(1356)$ are the phase-space residues of the vacuum amplitude in LTD, i.e. they are obtained from the residue on the corresponding causal propagators, at $\lambda_{456} \to 0$ and $\lambda_{1356}\to 0$, respectively. They represent the perturbative quantum fluctuations at one-loop with two final-state particles, and at tree-level with three final-state particles, respectively. For example, for the decay of a heavy scalar $\Phi$ into lighter scalars, the phase-space residues are 
\bea
&& \ad{3,\Phi}(456)  =  \frac{g_\Phi^{(1)} m_\Phi^2}{x_{12345}}
\left(L^{13\bar 4}_{23\bar 4 \bar 5, 125} +
L^{12\bar 5}_{23\bar 4\bar 5, 134} + 
L^{2345}_{134,125} \right)~, \nn \\ 
&& \ad{3,\Phi}(1356)  =  \frac{g_\Phi^{(1)} m_\Phi^2}{x_{135}}
\left(\frac{1}{\lambda_{13 \bar 4} \lambda_{134} \lambda_{1\bar 2 5} \lambda_{125}} \right)~,
\eea
where $g_\Phi^{(1)}$ encodes the interaction couplings, the factor $x_{i_1\cdots i_n} = \prod_{s=1}^n 2\qon{i_s}$ is the product of the corresponding on-shell energies and $L^i_{j,k} = \lambda_i^{-1} \left(\lambda_j^{-1} + \lambda_k^{-1} \right)$, with
\beq
\lambda_{i_1\cdots i_r \bar i_{r+1} \cdots \bar i_n} = \lambda_{i_1\cdots i_r}- \lambda_{i_{r+1} \cdots i_n}~.
\eeq

The integration measure is written in terms of two loop three-momenta
\beq
d\Phi_{\lb_1 \lb_2} = \prod_{j=1}^{2} \frac{d^3 \lb_j}{(2\pi)^3}~, 
\label{eq:integrationmeasure}
\eeq
i.e. six integration variables. However, each term in \Eq{eq:decayratescalarNLO} must satisfy energy conservation, which is encoded in
\beq
\ps{i_1\cdots i_n \bar a} =  2\pi \,  \delta(\lambda_{i_1\cdots i_n \bar a})~, 
\label{eq:dirac}
\eeq
and the decay is isotropic in the rest frame of the decaying particle. As a result, the decay rate depends on two independent integration variables, given the constraints imposed by the Dirac delta functions in \Eq{eq:dirac}; a polar angle, which is the angle between the two loop three-momenta, usually parametrized as $\cos{\theta} = 1-2v$, with $v \in [0,1]$, and the modulus of one of the loop three-momenta, mapped from $[0,\infty)$ to the finite interval $[0,1)$ in the numerical implementation. Explicitly, 
\beq
d\Phi_{\lb_1 \lb_2} \to \frac{1}{4\pi^4}\int_0^\infty \lb_1^2  d|\lb_1| \int_0^\infty \lb_2^2 d|\lb_2| \int_0^1 dv~,
\label{eq:measure}
\eeq
with 
\beq
\ps{45 \bar 6} =  2\pi \,  \delta\left( \sqrt{\lb_2^2+m^2} -\sqrt{s} \right)~,
\label{eq:delta1}
\eeq
and
\bea
\ps{135 \bar 6} &=&  2\pi \,  \delta\bigg( |\lb_1+\lb_2| \nn \\&+& \sqrt{\lb_1^2+m^2}+\sqrt{\lb_2^2+m^2} -\sqrt{s} \bigg)~,
\label{eq:delta2}
\eea
where $|\lb_1+\lb_2|  = \sqrt{\lb_1^2 +\lb_2^2
+2(1-2v) |\lb_1||\lb_2|}$ and $|\lb_i| = \sqrt{\lb_i^2}$. The two Dirac delta functions in \Eq{eq:delta1} and \Eq{eq:delta2} allow to express one of the integration variables in \Eq{eq:measure} in terms of the other two, which we solve analytically in the numerical implementation.

The most important feature of \Eq{eq:decayratescalarNLO} is that loop and tree-level contributions, i.e. contributions with different numbers of final-state particles, are treated simultaneously under the same integration measure. This property guarantees the local cancellation of singularities arising in the state-of-the-art approach, and thus avoids having to perform intermediate calculations in arbitrary spacetime dimensions~\cite{Bollini:1972ui,tHooft:1972tcz}. In addition, the resulting integrand is flatter than in other approaches, allowing for a much faster and more efficient numerical implementation.

%%%%%%%%%%%%%%%%%%%%%%%%%%%%%%%%%%%%%%%%%%%%%%%%%%%%%%%%%%
\section{Quantum integration of NLO decay rates}

\label{sec:quantum_impl}
%%%%%%%%%%%%%%%%%%%%%%%%%%%%%%%%%%%%%%%%%%%%%%%%%%%%%%%%%%

%%%%%%%%%%%%%%%%%%%%%%%%%%%%%%
\begin{figure}[t]
\begin{center}
\includegraphics[scale=0.6]{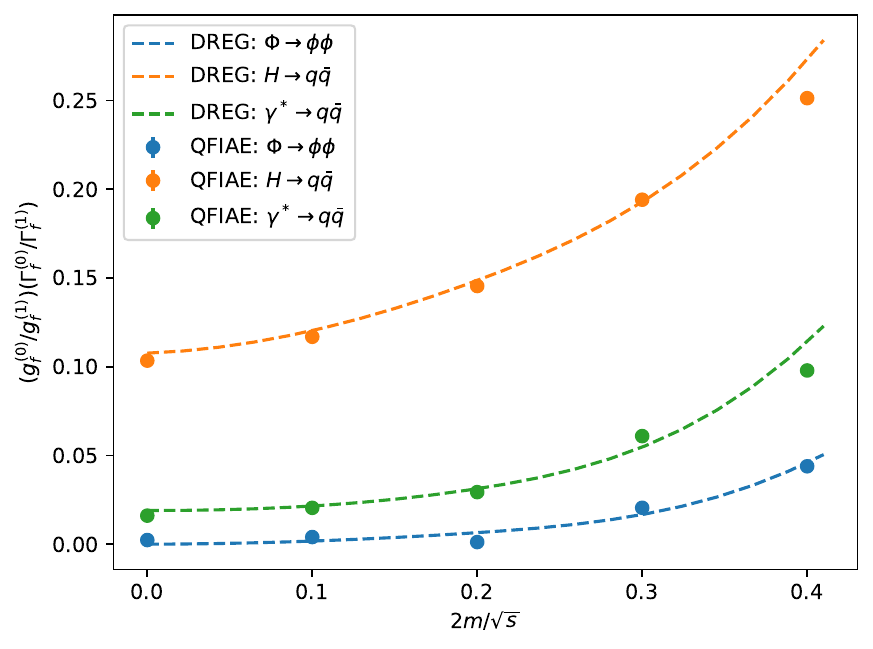}
\caption{Quantum-integrated decay rates in a quantum simulator for the three decay processes $H\to q\bar q (g)$, $\gamma^*\to q\bar q (g)$ and $\Phi\to \phi\phi(\phi)$ at NLO as a function of the final state mass, using QFIAE and LTD causal unitary. The dashed lines are the theoretical predictions in dimensional regularization. The parameters used in the quantum implementation are: $max\_steps=15000$, $step\_size=0.001$, $layers=n_{Fourier}=20$, $n_{qubits}=6$ for the QNN and $n_{qubits}=5$, $n_{shots}=10^3$,$\epsilon=0.01$, $\alpha=0.05$ for the IQAE  module. 
\label{fig:qint1}}
\end{center}
\end{figure}
%%%%%%%%%%%%%%%%%%%%%%%%%%%%%%

%%%%%%%%%%%%%%%%%%%%%%%%%%%%%%
\begin{figure}[t]
\begin{center}
\includegraphics[scale=0.6]{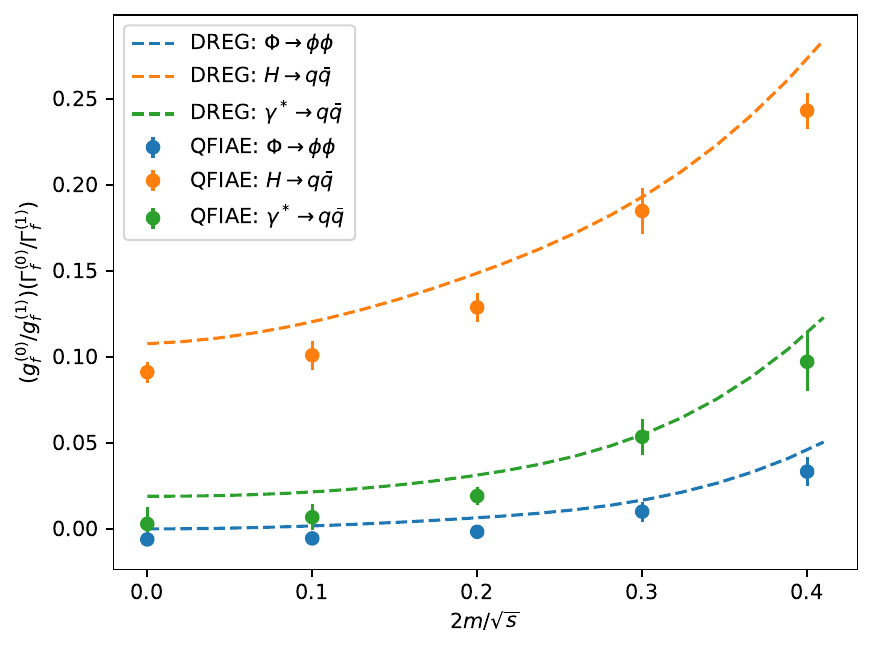}
\caption{Quantum-integrated decay rates in quantum simulator (hardware) for the QNN (IQAE) module of the QFIAE, for the three decay processes $H\to q\bar q (g)$, $\gamma^*\to q\bar q(g)$ and $\Phi\to \phi\phi(\phi)$ at NLO as a function of the final state mass, using QFIAE and LTD causal unitary. The dashed lines are the theoretical predictions in dimensional regularization. The parameters used in the quantum implementation are: $max\_steps=15000$, $step\_size=0.001$, $layers=n_{Fourier}=20$, $n_{qubits}=6$ for the QNN and $n_{qubits}=5$, $n_{shots}=10^3$,$\epsilon=0.01$, $\alpha=0.05$ for the IQAE  module. 
\label{fig:qint2}}
\end{center}
\end{figure}
%%%%%%%%%%%%%%%%%%%%%%%%%%%%%%

%%%%%%%%%%%%%%%%%%%%%%%%%%%%%%
\begin{figure}[t]
\begin{center}
\includegraphics[scale=0.37]{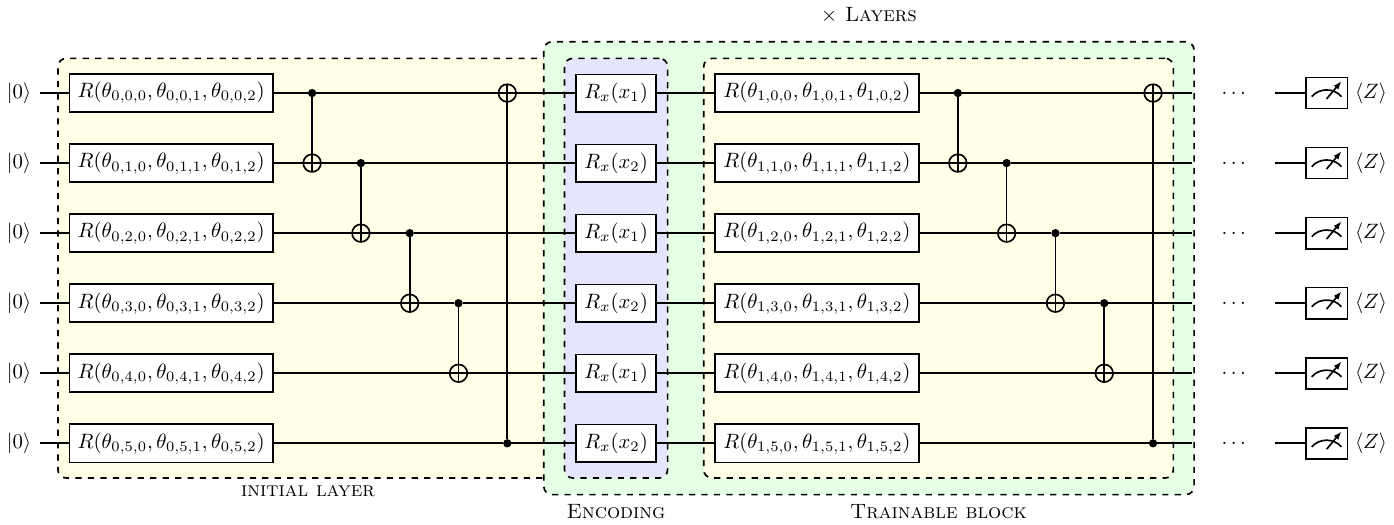}
\caption{Architecture of the QNN employed to fit a 2-dimensional function. 
\label{fig:qnn}}
\end{center}
\end{figure}
%%%%%%%%%%%%%%%%%%%%%%%%%%%%%%

In this section, we apply the quantum integration algorithm QFIAE to estimate the total decay rate at NLO of the decay processes presented in Section \ref{sec:ltd}. The main challenge of this quantum implementation is in making the QNN to fit well the differential decay rate function. To address this problem we present a QNN with a general-purpose Ansatz, see Fig. \ref{fig:qnn}, which contains enough entanglement and free parameters to permit a high expressibility that enables the correct solution of the regression problem.

%\bl{This could be confusing for the referee, because we start with the QNN, we move to IQAE, and then return to the QNN, 5 or 6 qubits is also confusing. I would first describe the QNN, and leave anything related to IQAE at the end} 
We utilize Pennylane~\cite{pennylane} to construct and train the QNN. The QNN architecture, which is displayed in Fig.~\ref{fig:qnn}, consists of a 6-qubit quantum circuit with a specific Ansatz repeated $n_{layers}$ times within the circuit. This Ansatz comprises two key components: a variational part with trainable parameters built using \texttt{qml.StronglyEntanglingLayers}, and an encoding block for two variables. Each variable is encoded three times in parallel across the 6 qubits using \texttt{qml.AngleEmbedding}. 
Building on previous studies of regression with variational quantum circuits \cite{deLejarza:2023IEEE, deLejarza:2024pgk}, the encoding has been performed using rotations $R_x$, and the measurements are performed on the Pauli-$Z$ basis, as depicted in Fig. \ref{fig:qnn}. Regarding the complexity of the QNN, each layer presents a quantum depth of 7, including one step of encoding, one of variational gates and five of entangling two-qubit gates. For the integrated decay rates shown in Figs.~\ref{fig:qint1} and ~\ref{fig:qint2}, 20 layers of the QNN architecture have been employed, which means that the total quantum depth of the variational quantum circuit is 140. To assess the feasibility of such variational circuit in current devices, we refer to two recent IonQ studies~\cite{ionqarticle,ionqnews}, where various quantum algorithms were evaluated using QED-C benchmarks. The findings show that our algorithm, which requires a quantum depth of 140 and a low qubit count ($\leq 6$), would achieve a high success probability on IonQ and Quantinuum devices.  

After making our QNN to accurately mimic the target function, we extract the Fourier series and feed the IQAE subroutine with it. For the IQAE module, we design a quantum circuit with a relatively low quantum depth and low number of qubits that opens up the possibility to be executed on current quantum computers. The IQAE module is implemented with \texttt{Qibo}~\cite{qibo_paper} on quantum simulators, Fig.~\ref{fig:qint1}, and with \texttt{Qiskit}~\cite{qiskit2024} on a real hardware, Fig.~\ref{fig:qint2}. In particular, the IQAE module is executed on the 27-qubit IBMQ superconducting device \textit{ibmq\_mumbai}. Only 5 qubits are needed for the implementation of the IQAE algorithm, which integrates the Fourier terms in sequential order.

We also implement error mitigation techniques to obtain the desirable results. In particular, to mitigate quantum noise during execution, we utilize a pulse-efficient transpilation technique~\cite{Earnest_2021}, which effectively reduces the number of two-qubit gates by leveraging the hardware-native cross-resonance interaction. Additionally, we apply the error suppression technique Dynamical Decoupling (DD) within the circuit execution and the error mitigation technique Zero Noise Extrapolation (ZNE) to the output, using the \texttt{Qiskit} Runtime Estimator primitive~\cite{estimator}.  

The hyperparameters used to train the QNN are $max\_steps$, which sets the number of iterations for the ADAM optimizer \cite{Kingma2014AdamAM}, $step\_size$, which represents the optimizer's learning rate, $layers$, which specifies the number of circuit layers, $n_{Fourier}$, indicating the number of Fourier coefficients we are truncating the Fourier representation of the circuit, and $n_{qubits}$, which defines the number of qubits used in the variational circuit. The IQAE parameters include $n_{qubits}$, specifying the qubits for the IQAE circuit, $n_{shots}$, for the number of measurement samples per circuit run, $\epsilon$, which controls the error tolerance of each individual integral, and $\alpha$, which defines the confidence interval for the integral results. 

The results presented in Figs.~\ref{fig:qint1} and~\ref{fig:qint2} show a relatively small deviation with respect to their corresponding analytical values in the standard dimensional regularization (DREG). In particular, in Fig.~\ref{fig:qint2}, one can notice that in comparison to Fig.~\ref{fig:qint1} there is a systematic deviation in the value of the integrals introduced by the hardware noise that is still not alleviated by the currently available error mitigation techniques applied. In Table \ref{table:errors}, we provide the explicit numerical results and uncertainties corresponding to Figs.~\ref{fig:qint1} and~\ref{fig:qint2}.

\begin{table}[t]
\begin{tabular}{ccccc}  \hline
Decay & $2m/\sqrt{s}$ & Hardware & Simulator &  DREG \\ \hline
%\multicolumn{2}{c}{\rt{Analytical}}& \rt{$-0.10073$} &  \rt{$-0.25547$}  \\ \hline
 $\Phi\to \phi\phi(\phi)$         & $0.0$ & $-0.0061(28)$ &  $0.0023(5)$ &  $0.0000$ \\ 
 %$\Phi$         
 & $0.1$ & $-0.0055(31)$ &  $0.0040(6)$ &  $0.0018$ \\
 %$\Phi$         
 & $0.2$ & $-0.0016(30)$ &  $0.0011(6)$ &  $0.0065$ \\
 %$\Phi$         
 & $0.3$ & $ 0.0101(56)$ &  $0.0205(11)$ &  $0.0167$ \\
 %$\Phi$         
 & $0.4$ & $ 0.0333(85)$ &  $0.0439(15)$ &  $0.0459$ \\ \hline
 $H\to q\bar q(g)$            & $0.0$ & $ 0.0911(61)$ &  $0.1034(13)$ &  $0.1077$ \\
 %$H$            
 & $0.1$ & $ 0.1009(83)$ &  $0.1169(14)$ &  $0.1204$ \\
 %$H$            
 & $0.2$ & $ 0.1288(85)$ &  $0.1455(14)$ &  $0.1486$ \\
 %$H$            
 & $0.3$ & $ 0.1847(135)$ &  $0.1941(20)$ &  $0.1928$ \\
 %$H$            
 & $0.4$ & $ 0.2431(104)$ &  $0.2513(30)$ &  $0.2730$ \\ \hline
 $\gamma^*\to q\bar q (g)$     & $0.0$ & $ 0.0029(96)$ &  $0.0161(14)$ &  $0.0190$ \\
 %$\gamma^*$     
 & $0.1$ & $ 0.0068(74)$ &  $0.0205(13)$ &  $0.0215$ \\
 %$\gamma^*$     
 & $0.2$ & $ 0.0191(50)$ &  $0.0293(13)$ &  $0.0313$ \\
 %$\gamma^*$     
 & $0.3$ & $ 0.0535(103)$ &  $0.0609(20)$ &  $0.0547$ \\
 %$\gamma^*$     
 & $0.4$ & $ 0.0971(171)$ &  $0.0979(30)$ &  $0.1140$ \\   \hline
\end{tabular}
  \caption{Quantum-integrated decay rates for the three decay processes $H\to q\bar q (g)$, $\gamma^*\to q\bar q(g)$ and $\Phi\to \phi\phi(\phi)$ at NLO as a function of the final state mass, using QFIAE and LTD causal unitary. The column ``HARDWARE'' contains the results obtained with the QNN on a quantum simulator and the IQAE on quantum hardware, whereas the column ``SIMULATOR'' contains the results obtained when both the QNN and the IQAE are executed on quantum simulators. The DREG column contains the exact analytic results at NLO accuracy. 
  }
  \label{table:errors}
%\vspace{-0.5cm}
\end{table}

Table \ref{table:errors} shows that the uncertainties from executing IQAE on quantum hardware are approximately an order of magnitude higher than those obtained on a quantum simulator. This difference is expected, as the inherent quantum noise on physical hardware adds to the statistical uncertainty in the IQAE method. Nevertheless, most of the values are in agreement within the uncertainty bands with the expected values, so we consider that the results are quite satisfactory, taking into account the current limitations of real quantum hardware.

%%%%%%%%%%%%%%%%%%%%%%%%%%%%%%%%%%%%%%%%%%%%%%%%%%%%%%%%%%%%%%%%%%%%%%%%%%%%%%%%%%%%%%%%%%%%%%%%%%%%%%%%%%%%%%%%%%%%%%%%%%
\section{Conclusions}
\label{sec:conclusions}
We have presented the first quantum computation of a total decay rate at second order in perturbative quantum field theory. Leveraging the loop-tree duality (LTD) framework, we have successfully combined loop and tree-level Feynman diagrams with a quantum algorithm on a quantum computer. This methodological advancement is significant from the high-energy physics perspective, as it allows us to integrate a real process with potential for quantum speedup. While we do not claim to have achieved quantum advantage in this work, our results lay the groundwork for future explorations in this direction.
From the perspective of quantum computing, our study marks a noteworthy achievement. By solving a relatively complicated regression problem using a Quantum Neural Network (QNN) on a realistic dataset, we found a good compromise between trainability and expressibility, a common challenge in quantum neural networks. Most of the results presented are in agreement with the expected values within the uncertainty bands. This demonstrates the potential of quantum computing to address complex, real-world problems and highlights the importance of continuing to push the boundaries of what quantum technology can achieve.
%Our findings underscore the critical importance of integrating quantum computing methodologies with high-energy physics applications. By tackling practical challenges and pushing the limits of current quantum computing capabilities, we can better understand the potential and limitations of quantum technologies and design better algorithms for future applications. 

%%%%%%%%%%%%%%%%%%%%%%%%%%%%%%%%%%%%%%%%%%%%%%%%%%%%%%%%%%%%%%%%%%%%%%%%%%%%%%%%%%%%%%%%%%%%%%%%%%%%%%%%%%%%%%%%%%
\begin{acknowledgements}
This work is supported by the Spanish Government - Agencia Estatal de Investigaci\'on (MCIN/AEI/10.13039/501100011033) Grants No. PID2020-114473GB-I00, No. PID2023-146220NB-I00 and No. CEX2023-001292-S, and Generalitat Valenciana Grants No. PROMETEO/2021/071 and ASFAE/2022/009 (Planes Complementarios
de I+D+i, NextGenerationEU). 
This work is also supported by the Ministry of Economic Affairs and Digital Transformation of the Spanish Government and NextGenerationEU through the Quantum Spain project and by the CSIC Interdisciplinary Thematic Platform on Quantum Technologies (PTI-QTEP+). DFRE and JML are supported by Generalitat Valenciana (CIGRIS/2022/145 and ACIF/2021/219).  MG is supported by CERN through the CERN QTI. Access to the IBM Quantum Services was obtained through the IBM Quantum Hub at CERN.
\end{acknowledgements}

%\appendix
%%%%%%%%%%%%%%%%%%%%%%%%%%%%%%%%%%%%%%%%%%%%%%%%%%%%%%%%%%%%%%%%%%%%%%%%%%%%%%%%%%%%%%%%%%%%%%%%%%%%%%%%%%%%%%%%%%%%%%%%%%

%%%%%%%%%%%%%%%%%%%%%%%%%%%%%%%%%%%%%%%%%%%%%%%%%%%%%%%%%%%%%%%%%%%%%%%%
%%%%%%%%%%%%%%%%%%%%%%%%%%%%%%%%%%%%%%%%%%%%%%%%%%%%%%%%%%%%%%%%%%%%%%%%
\bibliographystyle{JHEP}
\bibliography{main}

\end{document}